\documentclass[12pt,twocolumn]{elsart}
\usepackage{graphicx}
\usepackage{epstopdf}
\begin{document}
\begin{frontmatter}
\title{The BRAHMS Ring Imaging \v{C}erenkov Detector}
\date{January 31, 2002}
\author[BNL]{R. Debbe }
\author[NBI]{C. E. J{\o}rgensen}
\author[BNL]{ J. Olness }
\author[UBN]{ Z. Yin}

\address[BNL]{Brookhaven National Laboratory, Upton, New York 11973 USA}
\address[NBI]{CERN, PH-EP, CH-1211 Geneva 23, Switzerland}
\address[UBN]{Institute of Particle Physics, 
Huazhong (Central China) Normal University,
430079 Wuhan, China}

\begin{abstract}
A Ring Imaging \v{C}erenkov detector built for the BRAHMS experiment at the Brookhaven RHIC is 
described. This detector has a high index of refraction gas radiator. \v{C}erenkov light is
focused on a photo-multiplier based photon detector with a large spherical mirror. The 
combination of momentum and ring radius measurement provides particle identification from
2.5 GeV/c up to 35 GeV/c for pions and kaons  and well above 40 GeV/c for protons during runs that
had the radiator index of refraction set at $n-1=1700 \times 10^{-6}$.

\end{abstract}
\end{frontmatter}

\section{Introduction}

The BRAHMS (Broad RAnge Hadron Magnetic Spectrometers) Collaboration is one of the four heavy ion experiments 
running currently  at the Brookhaven RHIC. This experiment was designed to study particle
production from colliding systems as varied as p+p, d+Au and Au+Au in a wide rapidity range that extends from 0 to 4 units of rapidity (  maximum beam rapidity: 5.4 )  These measurements  are primarily based on the extraction of
 inclusive transverse momentum distributions of fully identified particles with moderate reach in transverse momentum. 

Particle identification at momenta lower than 
7 GeV/c is done via the time-of-flight technique and the more challenging task of identifying 
high momentum particles (up and above 30 GeV/c) is tackled with a photo-multiplier based Ring Imaging 
\v{C}erenkov detector (RICH) described in the present paper.  This detector is the last 
element of a two section, movable spectrometer called Forward Spectrometer (FS). This spectrometer was designed to measure high momentum charged particles at angles with respect to the colliding beams that range from 30 to  $2.3^{\circ}$. Tracking and momentum measurement is done with four dipole magnets, two TPCs in the 
first section of the spectrometer where occupancy can be high, and three drift chamber modules in the back section where lower occupancy and higher resolution provide precise measurement of momentum. 

\begin{figure}[htb]

{\includegraphics*[width=7cm]{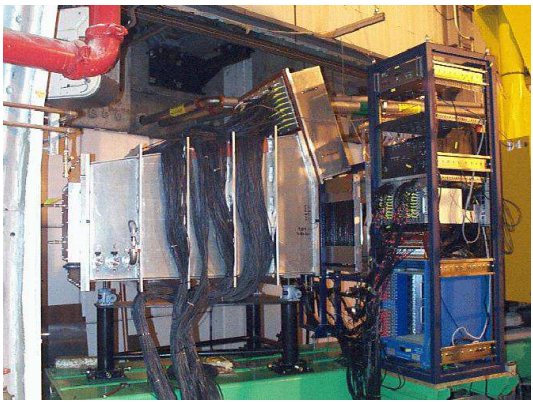}}

\caption{\label{fig:richPhoto} Photograph of the RICH installed at the end of the BRAHMS FS spectrometer. The interaction region is located some $\sim 20$ m to the right of the picture.}
\end{figure}

The BRAHMS RICH shown in Fig. \ref{fig:richPhoto} has a nominal radiator length of $150 cm$, and a $55 cm \times 70 cm$ spherical mirror with a radius $R = 3 m$. The mirror 
was manufactured as a slump formed glass substrate later polished to best commercial mirror surface quality. The polished surface was coated with Al and
a protective layer of $SiO$. The mirror is rotated by $8^{\circ}$  to keep the photon detector out of the spectrometer
acceptance. The photon detector is a photo-multiplier (PMT) based system described in section 3. This array of PMTs
is placed on the focal plane of the rotated mirror.  
More details about the BRAHMS experimental setup can be found in \cite{BRAHMSNIM}.

\section{Design and construction}

\v{C}erenkov light is a particular form of energy loss present  whenever a charged particle moves through the bulk of a
medium and its velocity is higher than the velocity of light in the medium $c/n$; this emitted light appears as
a conical ``retarded wake'' of light that follows the particle. 




The energy loss associated to \v{C}erenkov light emission is evaluated at distances much greater 
than the typical atomic  
lengths of the material and it  appears as radiated photons of
frequency $\omega$ whenever the velocity of the particle $\beta$ attains values such that the dielectric constant of the material
$\epsilon(\omega)$ and the particle velocity satisfy the relation  $\beta^{2}\epsilon(\omega) > 1$ see REf. \cite{Jackson}.  The energy radiated per unit length is
known as the Frank-Tamm relation:

$$ \frac{dE}{dx} = \frac{(ze)^2}{c^2} \int _{\epsilon(\omega)> \frac{1}{\beta^2}} \omega(1-\frac{1}{\beta^2\epsilon(\omega)} )d\omega $$     

The integrand between brackets is equal to $ \sin^2(\theta)$ where $\theta $ is the angle of the emitted photon with respect to the particle velocity. From these relations it is possible to 
extract an estimate of the amount of
light generated in a particular detector. The fact that \v{C}erenkov light emission does not depend on a microscopic description of the 
medium implies that the emission is coherent, and it is another quantum effect visible in macroscopic systems.  The index of refraction 
of the radiator medium written as:
 $ n = \sqrt{\epsilon(\omega)}$ can be measured for the frequencies of interest. A set of such measurements can be found  in \cite{Ypsilantis} and \cite{Ekelof}. 

The conical wave of \v{C}erenkov light is transformed into a plane wave by reflection
on a spherical mirror. If the angles of incidence to the mirror are small, 
all photons with the same angle of incidence are reflected 
onto a point on the focal plane, (located at R/2 where R is the radius of the mirror) and as \v{C}erenkov light is emitted 
uniformly in the azimuthal angle $\phi$ (defined with respect to the particle velocity), the
focus of \v{C}erenkov light reflected by an spherical mirror is a ring
centered around a point related to the angle of incidence of the particle with respect to the
axis of the mirror. Within the approximation of small angles, an spherical mirror focusing \v{C}erenkov light on a photon 
detector placed on its focal plane measures the charged particles angle of incidence and their 
velocity. 

The number of photon-electrons originating from a path of length L in the radiator volume is written as:

$$ N_{detected} =    N_{0} L \sin^2(\theta) $$

where $N_0$ (in units of $( cm)^{-1}$ ) contains all the detector parameters:

\begin{displaymath}
N_0=370  \varepsilon_m \tau_{quatz}f_{det}\int Q_{PMT}(\omega) d\omega
\end{displaymath}

where $\varepsilon_m$ is the reflectivity of the spherical mirror set as independent of frequency and equal 
to 90\%. $Q_{PMT}$ is the quantum 
efficiency of the photomultiplier photo cathode. The PMTs used in the BRAHMS RICH have borosilicate windows 
that set their quantum efficiency
$Q_{PMT}$  lower wave length cutoff  at 
250 nm and bialkali photocathodes with a 
maximum in  efficiency of 20\%. The calculated integral of $Q_{PMT}$ for these PMTs is equal to 0.33 eV. A fraction 
of the light, estimated to be equal to 9.7\%, is lost  
when the photons  traverse the 
 2.5 cm thick quartz window. The transmission of that window $\tau_{quartz}$ is thus set to 0.90. 
Finally the photon detector has dead
areas between PMTs that amount to an fiducial term $f_{det}$ equal to 0.7. The expected figure of merit of the
 BRAHMS RICH is then
equal to: $N_{0} = 69\ cm^{-1}$. 

The focused light doesn't form a perfect ring, chromatic and geometrical
aberrations will distort it. The position resolution of the photon detector will also contribute to the spread
of the photons away from a perfect circle. The angular or radial resolution of a Ring Imaging \v{C}erenkov
detector is  critical at high values of momentum where the bands of radius versus momentum for different 
particles begin to merge. The width of these bands sets the particle discrimination power. As mentioned above, the 
resolution of the detector will be written as the sum in quadrature of the chromatic 
aberrations and the single hit
radius resolution contribution:

\begin{displaymath}
 \left(\frac {\Delta r}{r}\right)^2  = \left(\frac {\Delta r} {r}\right)^2 _{chromatic}+\\
 \end{displaymath}
 \begin{displaymath}
 \left(\frac {\Delta r} {r}\right)^2_{geometric} + \left(\frac{\Delta r}{r}\right)^2_{detector} \
 \end{displaymath}


The contribution from  chromatic aberrations $(\Delta r/r)_{chromatic}$ is present because the index of 
refraction
in a dielectric depends on the frequency of the electromagnetic wave as $ n = \sqrt{\epsilon(\omega)}$.
Photons of different energy focus at 
different radii. For a particular detector sensitive to a finite band of wave 
lengths, the contribution of this distortion to the resolution of the detector can be simplified at high 
momentum values where $\beta \sim 1$ and the maximum angle of emission can be written as: \cite{Ekelof}

$$ \theta_{max} = \sqrt{1 - \frac{1}{n^2}} = \sqrt{\frac{(n+1)(n-1)}{n^2}}$$

the chromatic aberration can then be written as:

$$ \Delta \theta = \frac{\partial \theta}{\partial \lambda} \Delta \lambda = \frac{\partial \theta}{\partial n} \frac {\partial n } { \partial \lambda} \Delta \lambda =  \frac{\partial \theta}{\partial n} \Delta n $$

and

\begin{displaymath}
\frac{\Delta \theta}{\theta} = \frac{1}{\theta} \frac {\partial \theta}{\partial n}\Delta n \approx
\end{displaymath} 
\begin{displaymath}
\frac{1}{\theta_{max}} \frac {\partial \theta_{max}} {\partial n} \Delta n = \frac{\Delta n}{n(n^2-1)} 
\end{displaymath} 
  
The radius of a single focused photon has a fractional spread produced by chromatic aberration:

\begin{displaymath}
 \left(\frac{\Delta r} {r}\right)_{chrom} \approx \frac{\Delta \theta} {\theta} \approx \frac{\Delta n} {n(n^2 -1)} 
 \end{displaymath}

 $\Delta n$ in these relations is set by the dynamic range of the photon detector and can be evaluated 
from measured values. In the case of the BRAHMS RICH the photon detector is an array 
of multianode photo-multipliers with a FWHM quantum efficiency equal to 2.7 eV (from 250nm  to 517nm).
The radiator is a mixture of two heavy fluorocarbon gases $C_{4}F_{10}$ and $C_{5}F_{12}$ from 3M \cite{3M}.

The index of refraction of the gas mixture was calculated using measurements performed in liquid phase,
 \cite{Ypsilantis} and the Lorentz-Lorenz equation. Assuming an average index of refraction within the
dynamic range of the PMTs equal to $ n-1 = 1900 \times 10^{-6}$, the chromatic aberration contribution to the 
resolution in the measurement of a single photon is equal to: 

\begin{displaymath}
\left( \frac{\Delta r} {r}\right)_{chrom} = 0.0466 
\end{displaymath} 

 The four pixels of each R7600 photo-multiplier have square shapes with 1.1 cm on each side. For each of the measured photons, 
the error introduced in the ring radius measurement by the assumption that the photon intersects  the detector plane at the center 
of the pixel  is listed as:
 
 \begin{displaymath}
 \left( \frac{\Delta r} {r}\right)_{detector} = \frac{s}{r\sqrt{12}} 
 \end{displaymath}

 with s = 1.1 cm and maximum ring radius of 8.7 cm the contribution to the radius resolution  is equal to 0.036.
 
\begin{figure}[htb]
\begin{center}
{\includegraphics*[width=7cm]{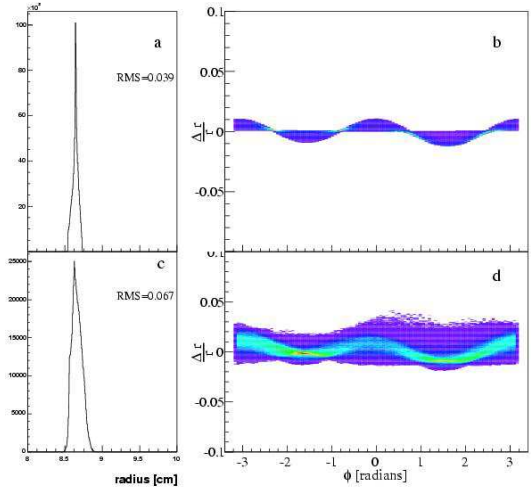}}
\end{center}
\caption{\label{fig:geoAberr} Geometrical aberrations from ray tracing in the actual geometry of the detector. Panels a and b correspond 
to the first set of pion moving all along the axis of the detector. Panels c and d were obtained with measured tracks that have different
angles of incidence as well as different entrance point locations, this set shows the strongest geometrical aberrations. For both sets
of pions, the left panel shows the distribution of radii, and the right ones display the relative deviation from a perfect circle as
a function of the azimuth angle. }
\end{figure}

 To estimate the overall effect of the  geometrical aberrations we have simulated the geometry of the detector and ray traced 
all \v{C}erenkov photons produced by charged pions with  fixed momentum set to 19 GeV/c. Two cases are considered, in the first 
one, a set of pions  move all along the axis of symmetry of the detector. The deviation from a perfect ring is shown in the  a 
and b  panels of Fig \ref{fig:geoAberr}.
 Panel a  shows the distribution of the distance between the photon intersection with the detector plane and a nominal center of 
the ring. The azimuthal dependence of the deviation from a perfect ring (for pions at 19 GeV/c and $n-1=1960$ ppm) is shown in panel b. 
These deviations are all the result of the 9 degree  rotation of the spherical mirror and the different positions of the emitted  
\v{C}erenkov 
photons along the track of the charged pion. The photons with the smallest deviations are the ones emitted close to the mirror. 
The overall effect of mirror rotation and photon emission  position is small and  transforms the rings into ellipses with the big 
axis along the x axis of the photon detector.
 
A second set of charged pions with the same fixed momentum was used to extract an overall effect from geometrical aberrations. 
This time the tracks belong to detected particles and give us the most realistic distributions in slopes and intersections with 
the radiator volume. The result of ray tracing the Cherenkov photons from this second set of pions is shown in panels c and d of 
Fig. \ref{fig:geoAberr}. The angles of incidence into the RICH are small, the most distorted rings are produced by charged particles 
that fly close to the edges of the radiator volume. We can thus quote a maximum value for geometrical aberrations as 
$(\Delta r/r)_{geometric} = 0.025$ even though the majority of the rings are contained in the bright band of panel d that 
corresponds to 0.7\% Once a particle is identified it is posible to correct the
geometrical aberrations in an iterative way. The present analysis does not include this correction.
     
A  prototype of this detector was tested in one of the experimental halls at the BNL AGS. These
studies aimed at developing a photo-multiplier based photon detector in collaboration with 
Hamamatsu Corp. From an earlier PMT version with 256 pixels that provided the first rings, 
the development work continued to reduce the amount of charge shared between neighboring
pixels. The second PMT version had $100 \times 1 cm^2$ pixel arranged in a 10X10 matrix. This device 
included an additional focusing stage between cathode and first dynode reducing the charge 
spread to 10\% at the center of the next pixel \cite{FirstNIM}, but had poor pulse height resolution, 
and was deemed too difficult  to manufacture.
Finally we tested the R5900 four pixel PMT mounted in a compact metallic can. In order to
achieve close packing af these tubes Hamamtsu produced the R7600 03 M4F that was selected 
to be used in this detector. Results obtained with the prototype can be found  in Refs. \cite{FirstNIM} and \cite{Prototype}.

\section{The photon detector}

 The photon detector is an array of 80 four-pixel photomultipliers R7600-03-M4F 
\cite{Hamamatsu}. Each PMT  has a single photocathode plane evaporated on a borosilicate 
window. Eleven dynodes amplify the electron extracted from the cathode and the total charge is
collected in four independent anode plates. A focusing stage is placed between cathode and dynodes 
to reduced the charge shared between anodes. The complete system is encased in a flange-less
metallic can held at the cathode voltage.

Close packing of these PMTs was achieved by Hamamatsu engineers with the design of a matrix of
5 x 4 biasing bases and sockets. Each matrix has two high voltage connections such that a 
single channel of a HV power supply  can bias ten PMTs. The photo-multipliers  placed
in these matrices were selected to have similar gains and pulse height resolutions.
The signal from each anode is routed out of each matrix through RG17 cables. 
The complete photon detector consists of four PMT matrices, and 70\% of the total area is covered by active pixels.

The output of these PMTs is fast and has sufficient pulse height resolution, (in average
the single photo-electron peak appears as a shoulder of the distribution) but their gain is
insufficient to send the signal through some 100 meters of cable (RG58) to be integrated in 
an ADC with 50 fC coulomb resolution. This shortcoming of the PMT was compensated by the
addition of one stage of amplification mounted right on the matrices of PMT bases. 
Fast charge amplifiers were AC coupled to the anode output through impedance matching 
$24 \Omega$ resistors to ground and a $0.1 \mu f$ capacitor. Fig. 3 shows one of the bias array with
the five amplifier cards connected to it. 

The response of the R7600 PMTs was studied with a well collimated LED positioned in front
of the photo-cathode with a two axis stepping motor system set to move in steps of 1 mm. The amount of light produced 
with green LED is constant during the duration of the scan of one photo-multiplier and
can thus be used to study the position dependence of the response of a particular pixel.
As the LED is 1 mm into
the neighboring pixel it has 20\% of the average pulse height in the center. Two mm further
away that fraction drops to $ \sim 6\% $ and is negligible at the center of the next pixel.

The 320 signal cables are connected to a LeCroy 1885F Fastbus ADC after suitable delay to
allow for a trigger definition for each RHIC collision. The charge on each channel was 
integrated during a gate that lasted 120 ns. 

\begin{figure}[htb]
\begin{center}
{\includegraphics*[width=7cm]{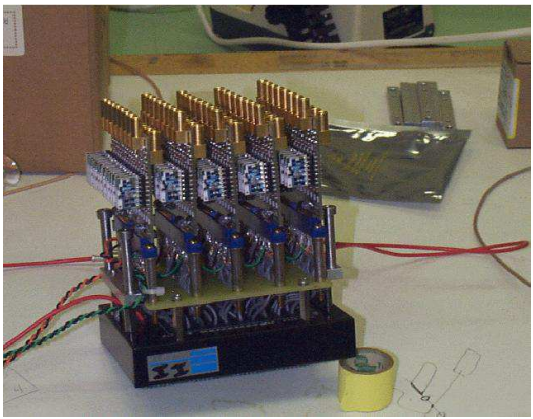}}
\end{center}
\caption{Photograph of 5 amplifier cards mounted on one of the 5X4 PMT base matrices.}
\end{figure}

\section{Filling procedure.}

Work performed on the prototype detector proved that the only filling procedure that 
guaranties the highest concentration of the gas mixture in the radiator volume is the one
that starts by evacuating the volume. The drawback of this method is the need of a
vessel designed  to withstand a full atmosphere pressure differential. 
Once the radiator volume was evacuated, a boil off of $C_5F_{12}$ was introduced till the 
pressure reached 392 Torr (or 41.3\% of the final mixture pressure of 1.25 atmospheres).
After that, $C_4 F_{10} $ is sent into the radiator volume till the final mixture pressure is reached. Once the filling is done, 
a small sample of the gas was used to measure the index 
of refraction by counting fringes in a Young interferometer as one of the split beams of laser 
light goes through a volume that is filling with the gas sample, and the other passes 
through an equal length of vacuum.

\begin{figure}[ht]
\begin{center}
{\includegraphics*[width=7cm]{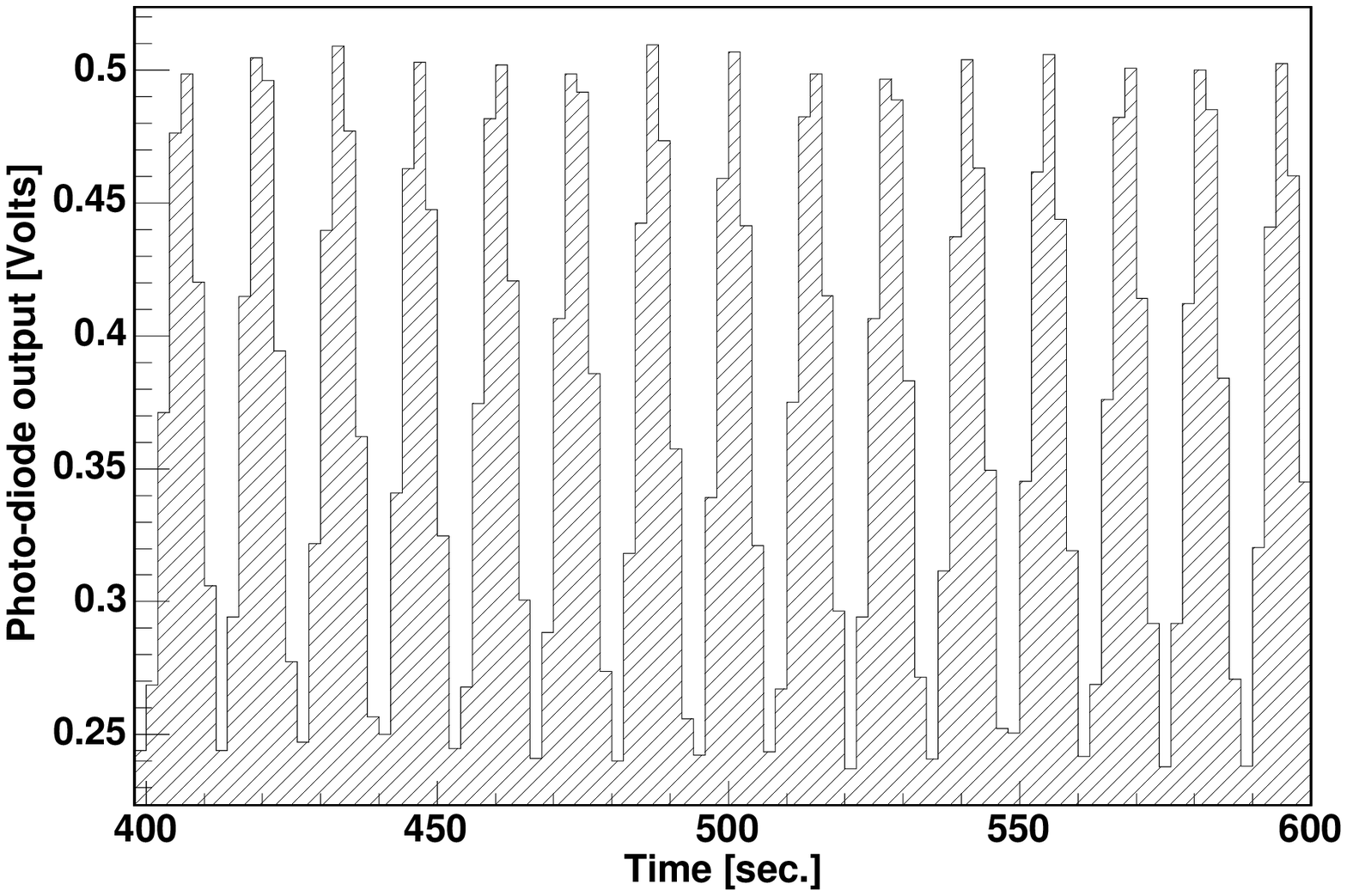}}
\end{center}
\caption{\label{fig:fringes} Interference fringes detected with a PIN diode while a sample of the radiator
gas is brought into one of the cavities of the interferometer while the other is kept at vacuum.}
\end{figure}

Fig. \ref{fig:fringes} displays a portion of the fringes detected with a
PIN diode to convey their good definition that makes possible to make measurement of the index of refraction with a resolution of 
one part in a million.

The highest index of refraction achieved with this mixture at 1.25 atmospheres was $ n-1 = 2029 \times 10^{-6}$. 
Later, the focus of the collaboration shifted to studies at higher $p_{T} $ values and the gas mixture and operating 
pressure were changed to reduce the index of refraction to lower values ($n-1 = 1600 \times 10^{-6}$).  

\section{Data analysis}

Tracks reconstructed with the FS spectrometer tracking system 
are used
to find the nominal center of rings. In case there are several tracks in the event, the loop
over tracks is ordered such that it starts with the one with the highest momentum. 
Once a ring center is defined, the distance from pixel
center to ring center is calculated for all pixels that have charge above pedestal. 
 The radius of a ring candidate is set as the average of those distances. 
The set of pixels is accepted as a ring if their number exceeds a minimum set by default 
to be equal to 4 and is at least 40\% of an estimated number of photo-electrons with a
figure of merit set low as $N_0 = 55$.

 Pixels that were included in the
histogram are marked and will not be used in the search for another ring in events where
there is more than one track in the RICH. A more detailed description of this algorithm can be found in \cite{ClausThesis}.

\begin{figure}[ht]
\begin{center}
{\includegraphics*[width=7cm]{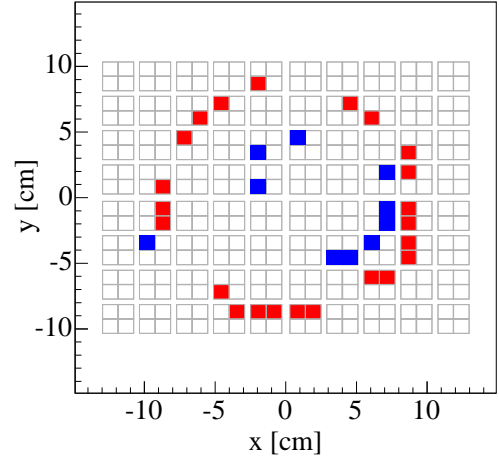}}
\end{center}
\caption{\label{fig:ring}Two rings of \v{C}erenkov light focused on the photon detector. The big ring (red pixels online) has been produced by a 20 GeV/c pion. The smaller ring (blue pixels online) was produced by a 17 GeV/c proton. }
\end{figure}

\section{Performance}

\begin{figure}[htb]
\begin{center}
{\includegraphics*[width=8cm]{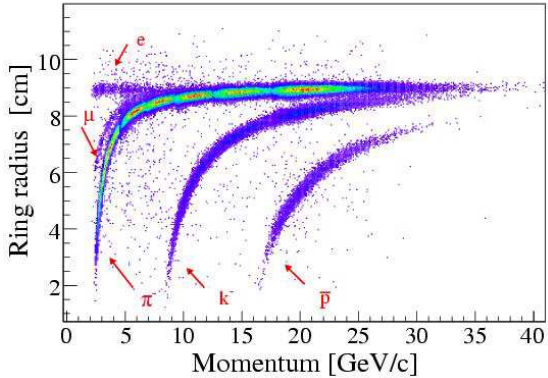}}
\end{center}
\caption{\label{fig:radiusVsP}The radius of the \v{C}erenkov rings produced by negative particles at 4 degrees with respect to the beams
in p+p collisions at $\sqrt{s} = 200$ GeV/c, as a function of their momentum multiplied by the charge. Different magnetic field settings
of the FS spectrometer are included in this figure. No effort is made to normalize the different data samples. }
\end{figure}

Fig. \ref{fig:radiusVsP} is a composite of five field settings of the FS spectrometer, no effort was made to normalize the yields, 
the main purpose of this figure is to show the remarkable momentum range of this detector; it can identify electrons, muons and pions 
at low momentum,  kaons are well separated from pions (at the three standard deviation level) up to $\sim 25$ GeV/c.
The index of refraction of the radiator gas throughout the runs was equal to $n-1 = 1690 \times 10^{-6}$ and the spectrometer was placed 
at four degrees with respect to the beam line.  The rings with "saturated"
radii extracted from runs where the index of refraction was set to $ n-1=1560 \times 10^{-6}$ 
have an average of $38\pm 9$ photo-electrons. The measured figure of merit of this detector is thus $N_{0} = 81 \pm  16 cm^{-1}$. 

\begin{figure}[ht]
\begin{center}
{\includegraphics*[width=7cm]{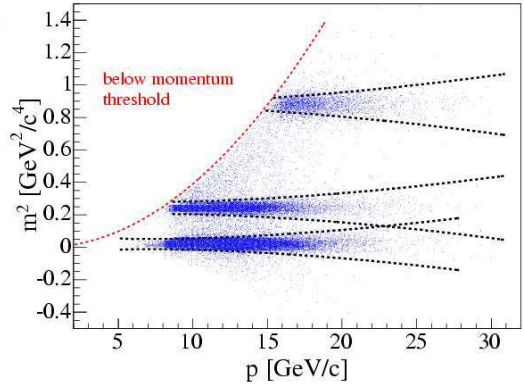}}
\end{center}
\caption{\label{fig:mass2}  Mass-squared as a function of momentum. The dashed curves show the 
$\pm 2\sigma_{m^2}$ cut used by the mass based particle identification method. The red dashed curve
shows the threshold for \v{C}erenkov light emission as function of momentum.}
\end{figure}

Particle identification with the RICH detector can be done with two independent and consistent
methods, the first one is based on the difference between measured ring radii and the expected
radius of a ring produced by a candidate particle. If such difference falls within a set number of
standard deviations the measured particle identity is set to be the one of the candidate particle.
This method requires the correct value of the index of refraction of the radiator gas extracted previously from the data and stored in a run information database. This method includes tools described 
in section 6.2  to
handle high momentum particles  whenever the pion and kaon band start to overlap.  The second 
particle identification method is based on the calculated mass using the momentum of the particle, 
the radius of the \v{C}erenkov ring and the index of refraction of the radiator. The resolution of this
calculated mass is momentum dependent and has contributions from the momentum resolution as well
as the radius resolution. Fig. \ref{fig:mass2} shows the distribution of mass squared as a function
of momentum. The particle identification is done in this particular case with a $\pm 2\sigma_{m^2}$ 
cut indicated with dashed curves, where $\sigma_{m^2}$ is the standard deviation of the mass square 
distribution.

\begin{figure}[htb]
\begin{center}
{\includegraphics*[width=8cm]{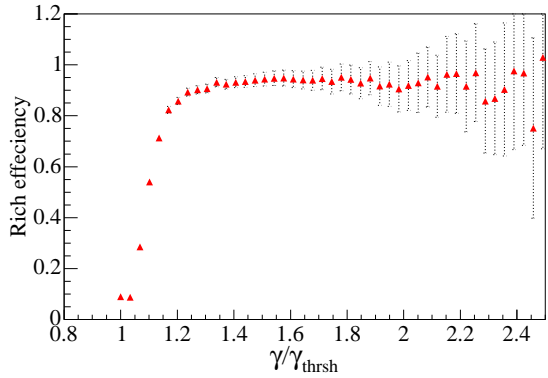}}
\end{center}
\caption{\label{fig:richEffic} Efficiency near threshold calculated using protons. The horizontal axis is a normalized value of the $\gamma$ of the particle }
\end{figure}

Monte Carlo simulations of the BRAHMS spectrometer, together with information extracted from data, show the high efficiency of this detector; Fig. \ref{fig:richEffic} displays the efficiency as function of the ratio $\gamma / \gamma_{thrsh}$ where $\gamma_{thrsh}$ is the value of the $\gamma$ factor at the particle threshold. The efficiency values shown in this figure were obtained with protons.  The simulations show a $\sim 4 $\% inefficiency due to interactions with material at the entrance of the RICH. 

\subsection{Relative radius resolution}

\begin{figure}[htb]
\begin{center}
{\includegraphics*[width=8cm]{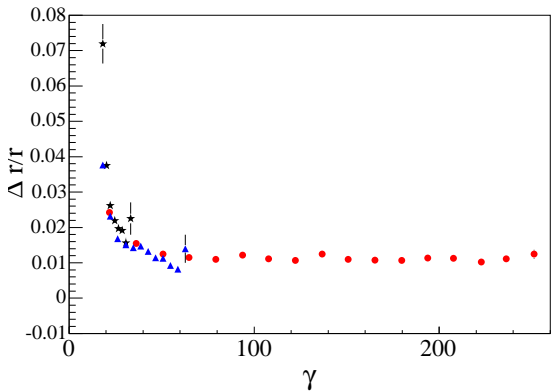}}
\end{center}
\caption{\label{fig:resolution}A fit to the width of the pion band.}
\end{figure}

The particle discrimination power of this detector is set by the relative radius resolution at each momentum value. Fig. \ref{fig:resolution} shows the standard deviation obtained from gaussian fits to  the distributions of the ratio $(r_{measured} - r_{calc})/r_{measured}$, where $r_{calc} $ is the expected radius for rings produced by pions, kaons or protons.

The horizontal axis displays the velocity of the particles shown as their $\gamma$ factor. Pions are shown with red filled circles and above $\gamma \sim 40$ have a constant relative radius resolution with a value as low as 1.2\%, kaons shown with blue filled triangles and the anti-protons shown with filled star symbols have a worsening resolutions  as their momentum diminishes.

\subsection{Particle identification at high momentum}

\begin{figure}[htb]
\begin{center}
{\includegraphics*[width=8cm]{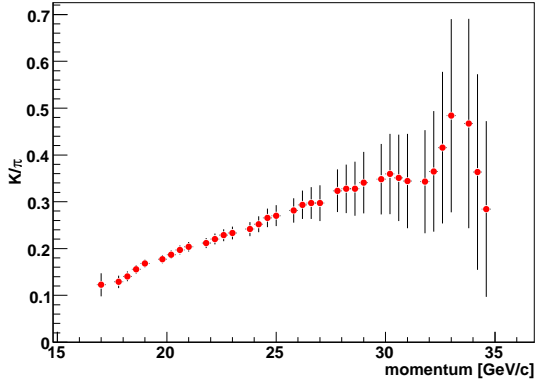}}
\end{center}
\caption{\label{fig:ratioKpi}The abundance of positive kaons with respect to pions in p+p collisions at high rapidity ($y\sim 3$) obtained  from fits to projections of narrow momentum bands onto the radius axis.}
\end{figure}

Fig. \ref{fig:radiusVsP} shows that kaons can be separated from pions with simple cuts of the radius of the rings for momenta smaller than $\sim 25 GeV/c$, and that protons are well separated beyond the momentum range of the figure. In order to extend the separation of kaons and pions to higher values of momentum it is necessary to parametrize the uncorrected abundance of kaons and pions in small momentum bins in order to allocate probabilities or weights to events where the kaon band has started merging with the pion band. That parametrization was obtained at each full field spectrometer setting by fitting projections of narrow momentum bands (500 MeV/c) onto  the radius axis. The functional form used for these fits was the sum of two gaussians with equal widths, the free parameters of the fit were the normalizations, centroids and the common width. Fig. \ref{fig:ratioKpi} shows the results of those fits.

\begin{figure}[htb]
\begin{center}
{\includegraphics*[width=8cm]{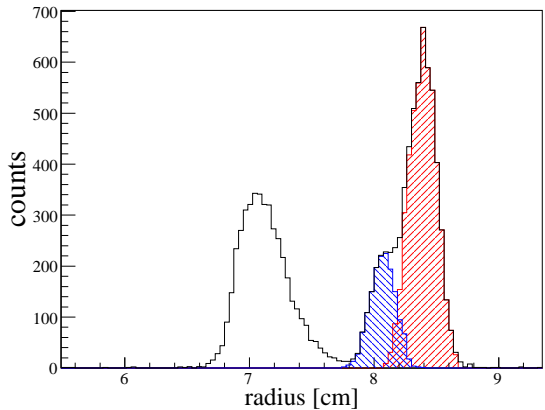}}
\end{center}
\caption{\label{fig:resultPID}This figure illustrates the fact that particle discrimination  between kaons 
and pions is still possible even though their radius versus momentum bands start to overlap, the assignment of probabilities is described in the text. }
\end{figure}

Once the abundance of kaons with respect to pions is know for a particular full field spectrometer it is possible to assign probabilities to events that lie in the overlap of the kaon and pion bands. Fig. \ref{fig:resultPID} shows the separation between kaons (blue hatched histogram  centered around 8.1 cm) and the more abundant pions shown as the red histogram centered at 8.4 cm) in a wide momentum bin (from 30 to 44.6 GeV/c). Protons are also present in this figure but their distribution is not gaussian because at these momenta, the rings radii are still changing fast as function of momentum.

\section{Conclusion}

The BRAHMS RICH described here has performed very well throughout the six years of data collection at RHIC. Its extended particle identification range has been instrumental in the study of particle production at high rapidity in several nuclear system that include p+p, d+Au and several heavy-ion systems. The high resolution of the radius measurement together with the simplicity of the photo-multiplier based photon detector have made this detector the most important tool among the other particle identification counters in the BRAHMS setup.

\section{Acknowledgments}

We are greatful to the members of the BRAHMS Collaboration whose participation made this work possible, special thanks to F. Videb\ae k and Ch. Chasman for their help and valuable suggestions.
This work was supported by 
the Office of Nuclear Physics of the U.S. Department of Energy, 
the Danish Natural Science Research Council, 
the Research Council of Norway, 
the Polish State Committee for Scientific Research (KBN) 
and the Romanian Ministry of Research.

\end{document}